# Fluidic Shaping and *in-situ* Measurement of Liquid Lenses in Microgravity

Omer Luria [a], Mor Elgarisi [a], Valeri Frumkin [a,§], Alexey Razin [a], Jonathan Ericson [a], Khaled Gommed [a], Daniel Widerker [a], Israel Gabay [a], Ruslan Belikov [b], Jay Bookbinder [b], Edward Balaban [b,*] and Moran Bercovici [a,*]

[a] Faculty of Mechanical Engineering, Technion – Israel Institute of Technology, Haifa, Israel

[b] NASA Ames Research Center, Moffett Blvd., Moffett Field, CA, USA

[§] Current affiliation: Department of Mathematics, Massachusetts Institute of Technology, Cambridge, MA, USA

[*] Corresponding authors: edward.balaban@nasa.gov (E.B), and mberco@technion.ac.il (M.B)

**Abstract**

In the absence of gravity, surface tension dominates over the behavior of liquids. While this often poses a challenge in adapting Earth-based technologies to space, it can also provide an opportunity for novel technologies that utilize its advantages. In particular, surface tension drives a liquid body to a constant-mean-curvature shape with extremely smooth surfaces, properties which are highly beneficial for optical components. We here present the design, implementation and analysis of parabolic flight experiments demonstrating the creation and in-situ measurement of optical lenses made entirely by shaping liquids in microgravity. We provide details of the two experimental systems designed to inject the precise amount of liquid within the short microgravity timeframe provided in a parabolic flight, while also measuring the resulting lens' characteristics in real-time using both resolution target-imaging and a Shack-Hartmann wavefront sensing. We successfully created more than 20 liquid lenses during the flights. We also present video recordings of the process, from the lenses' creation during microgravity and up until their collapse upon return to gravity. To the best of our knowledge, this is the first time that a purely liquid lens has been created in microgravity, which demonstrates the feasibility of creating and utilizing liquid-based optics in space.

## 1. Introduction

Optical components are used in a wide range of space applications, including imaging, spectroscopy, communications, and solar concentration[1–3]. Fabrication of optical components is traditionally based on mechanical processes such as grinding and polishing that produce significant waste and rely on heavy machinery[4–8]. Such processes are therefore not suitable for implementation in space. Although 3D printing overcomes many of these limitations and has been successfully utilized in space, it cannot yet produce optical-grade components due to the resulting low surface quality[9–11]. As a result, currently all optical components, from small lenses to large-scale telescope mirrors, are fabricated entirely on Earth and launched into space[5,12]. The ability to manufacture optical components in space could greatly benefit long-duration missions (e.g. to Mars) that must be self-sufficient, as well as open the door to creation of large-scale telescopes, breaking away from the current limitations imposed by the launch process[13–15].

The use of liquids as a method for creating space optics has been of interest for decades. One prominent example is the 'spinning liquid telescope' which subjects reflective liquids to centrifugal forces to create a parabolic mirror. Such telescopes have been demonstrated on Earth[16–18] and have been suggested as a technology for a lunar-based telescope[19]. However, this approach is not suitable for use in microgravity conditions because, in addition to spinning, it also requires a uniform body force along the optical axis to form the parabolic shape. Moreover, implementation of such telescopes requires high mechanical precision, dynamic stability, and continuous energy consumption to sustain their shapes[17]. Another use of liquids in optics are the so called 'liquid lenses', wherein an optical liquid is confined between two elastic membranes, and the change in focal length is achieved by mechanical or electrical actuation of the liquid[20–22]. These 'liquid lenses' provide rapid changes in focal length, making them attractive for machine vision applications[23]. Through parabolic flight experiments, Newman and Stephens tested the effect of microgravity on their performance[24]. While this technology is likely useful in space, it still relies on high-quality solid components (e.g. the membrane) that must be prepared on earth, limiting its use and scale.

Recently, a new method for additive manufacturing of optical components was reported[25]. The method, termed 'Fluidic Shaping', uses surface tension under neutral (or near neutral) buoyancy conditions to shape a volume of liquid into useful optical components by contacting it with a rigid bounding frame that serves as a spatial constraint. Neutral buoyancy is achieved using an insoluble immersion liquid that has equal density to that of the optical liquid. Both spherical and aspherical lenses, as well as freeform components, can be produced, with the specific topography controlled by the shape of the bounding frame and the level of deviation from neutral buoyancy[26]. In perfect neutral buoyancy, surface tension dominates completely and results in scale-invariant constant-mean-curvature surfaces. For a circular flat boundary (a simple ring) the liquid volume takes a spherical cap shape[25], with a curvature (positive or negative) determined by the volume of the optical liquid. The component can either remain liquid and thus allow for dynamic control of its curvature by adding or aspirating liquid, or be solidified (e.g., polymerized) to form a solid object.

Owing to its simplicity and inherent compatibility with microgravity, Fluidic Shaping has the potential to serve as a method for in-space manufacturing of high-quality optical components. Moreover, microgravity eliminates the need for a matching immersion liquid. This further simplifies the process relative to its Earth-based implementation by making it independent of the liquid's absolute density or solubility. Due to the scale invariance of the method under perfect microgravity conditions, optical components of any size can be theoretically produced, while maintaining the same surface quality.

In this work we present, for the first time, the use of Fluidic Shaping in creation of liquid optical lenses in microgravity without an immersion liquid, in a series of parabolic flight tests. We show the feasibility of the method by injecting a lens liquid into a bounding frame during microgravity, resulting in 20 successful deployments of liquid lenses suspended in air. We present the performance of resulting lenses obtained by *in-situ* resolution-target imaging and Shack-Hartmann wavefront sensing, within the short microgravity time. The results are accompanied by a detailed design of the hardware, a discussion of design considerations, and guidelines for the construction of such setups.

While our demonstration here was limited to a relatively small scale — due to practical constraints of the on-flight experimental environment — it clearly demonstrates the feasibility of creating liquid optics in microgravity. This potentially opens the door to in-space fabrication of optical components and provides a path toward space telescopes that are based on deployment of liquids in space on scales that cannot be reached using today's technologies.

## 2. Experimental Hardware

### 2.1. Overview and lens deployment approach

The experiment was executed on a reduced-gravity aircraft (Boeing 727-227F, Zero-G Corporation, FL, USA) performing parabolic maneuvers, each providing a microgravity time window of roughly 15 seconds. Figure 1a shows a general schematic of the hardware structure. Liquid is pushed into a bounding frame to create the lens under test (LUT). A g-sensor (accelerometer) registers the proper acceleration (i.e., the acceleration relative to free-fall), and a GoPro side camera video records the entire process. Two types of experimental setups were used, which differ only in the way that the LUT was characterized *in-situ* - one by capturing an image of a test target through the LUT using a DSLR camera ('DSLR setup', shown in Figure 1a), and the other by measuring the effect of the lens on incident light using a Shack-Hartmann Wavefront Sensor ('SHWS setup', not shown in the figure).

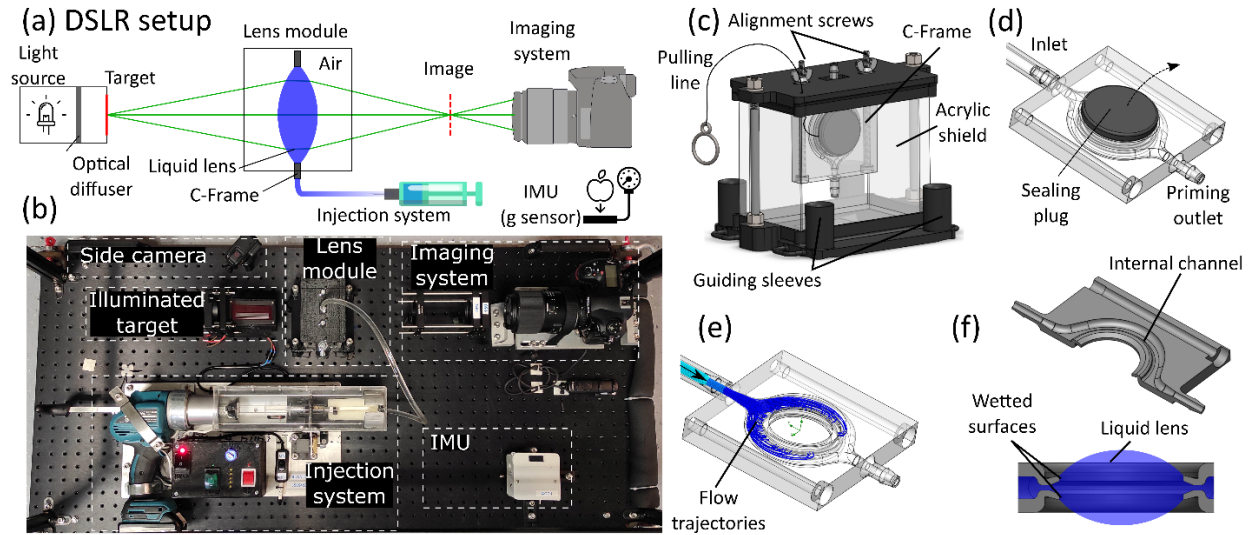

*Figure 1: (a) Schematic and (b) photo of the experimental hardware, showing the central elements. In microgravity, the lens is formed by injecting liquid into the C-Frame contained within the aquarium module. In parallel, an optical system characterizes the resulting lens by imaging through it, a side camera captures a video of the process, and a g-sensor records the proper acceleration. The entire system is mounted on an optical breadboard for structural and optical robustness. (c) The lens module, showing the C-Frame enclosed in the acrylic shield with two screws used for alignment. The module is aligned by sliding its guiding sleeves along four vertical rods on the breadboard, and the pulling line is used to pull the plug before injecting the liquid through the primed channel. (d) The hole at the center of the frame is initially blocked by a plug that prevents the liquid in the channel from leaking out. (e) After pulling the plug and activating the injection system, liquid flows through the channel and subsequently fills the hole. (f) Cross-section of the C-Frame, showing the inner channel and the wetted surfaces which serve as the bounding frame for the resulting liquid lens.*

Figure 1b shows a top-view photo of one of the DSLR setup. A syringe was used to inject the optical liquid (silicone oil with several kinematic viscosities: 200, 1000 and 5000 cSt) into a bounding frame enclosed within an acrylic shield. We automated the injection using a customized, high-torque pump based on an electrical caulking gun powered by an 18v Li-ion battery (Makita DCG180). We controlled the DC motor of the gun directly with a microcontroller (Arduino Mega 2560) and an H-bridge driver (HIP4081AIVZ, Shenzhen LC Technology co.). We configured the injection system to accurately inject the required amount of viscous liquid within the first few seconds of each microgravity period. The injection system worked in open loop, based on pre-calibration of the flow rate for each viscosity as a function of the pulsed width modulation (PWM) signal. We measured the proper acceleration of the setup using an inertial measurement unit (IMU, SparkFun MPU-6050), interfaced by a microcontroller (Arduino Nano) powered by a set of AA batteries.

Figure 1c-f shows the heart of each setup - the "C-Frame", which is a 3D printed component with a 25 mm diameter hole at its center, serving as the bounding frame. The perimeter of the hole is an internal

circumferential channel with a C-shaped cross-section, whose opening is towards the center of the hole. The channel's inlet is connected via a 1.5 mm thick polyurethane tube (rated to a working pressure of 10 bar) to a syringe filled with the optical liquid. On the ground, before the experiment, we first sealed the C-Frame hole with a rubber plug and primed the channel by running liquid through it while pushing all the air through the outlet. Once all the air was removed and no bubbles were visible, we sealed the outlet with a clamp. To deploy the lens during the experiment in microgravity, we first pulled the sealing plug out of the hole. We then manually turned on the pump to push liquid from the syringe through the inlet. The liquid flowed radially into the hole until all liquid fronts met and a continuous volume of liquid was formed, creating a bi-convex spherical liquid lens. In the remaining microgravity time (typically 10 s) the formed lens is measured by the imaging system. Upon return to gravity, the liquid lens collapsed to the bottom of the acrylic shield.

Each such lens module (C-Frame within acrylic shield) was used a single time. Between sets of parabolas, the used modules were removed, and new ones were installed to repeat the experiment with a clean assembly. The C-Frame was connected to the module using two alignment screws that allowed lateral displacement for optical alignment during pre-flight assembly. The guiding sleeves at the bottom of each module allowed quick and easy mounting on top of the breadboard platform during the flight, while maintaining the optical alignment.

## 2.2. Optical measurement design

Figure 2a-b shows the optical designs used for the DSLR and SHWS imaging systems, respectively. In the DSLR design shown in (a), the illuminated target is a transmissive resolution test-chart made of a negative chrome mask etched on a glass substrate. The target is back-illuminated by a 530 nm LED board (CREE C503-GCN) with two layers of parchment paper acting as a diffuser, creating an extended source with a highly uniform radiance. The target is imaged through the LUT and creates an intermediate real image that is then magnified by the condenser (Thorlabs LB1374-A, f=150mm, d=50mm) and imaged through the objective (Tamron 90mm f/2.8 with a 12mm macro extension tube) to create the final image on the DSLR sensor (Nikon D850). To allow greater flexibility in finding the focus during the experiment, the target surface was fixed at an angle of 7.6º from the optical axis and the objective was manually moved against the sensor in real-time. The optical train was designed such that the resulting aberrations could be primarily attributed to the LUT, as shown in figure S1 in the Supplementary Information (SI).

In the SHWS design shown in (b), a point source (Thorlabs M530F2 530 nm fiber-coupled LED) is positioned at the back focal point of a collimator lens (Thorlabs LBF254-150-A) to create a planar wavefront. This beam then passes through the LUT and the resulting aberrated wavefront is measured by a Shack-Hartmann wavefront sensor (SHWS, Thorlabs WFS40-7AR).

Since the nominal LUT diameter is fixed to be the hole in the C-Frame (25 mm) and the refractive index of the silicone oil is provided by the manufacturer (1.403), uncertainties in volume can be directly translated to variations in the focal length. Figure 2c shows the relation between the volume of the LUT (modeled as two positive spherical caps and a disk) and its focal length, computed using the lens makers' equation for thick lenses[27,28]. The derivative of this curve is also shown, indicating the sensitivity of the focal length to the volume at different regions.

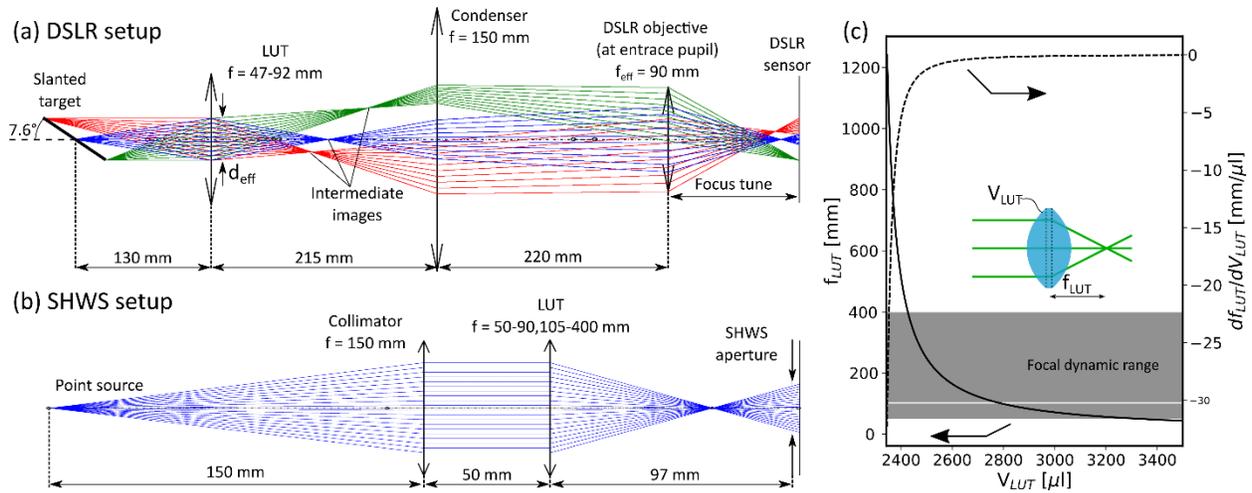

*Figure 2: The optical design of the two imaging system types, illustrated by paraxial ray tracing. (a) The DSLR setup design, where a resolution target is imaged through the lens under test (LUT) to form an intermediate real image, which is then magnified by the condenser and imaged through the objective to form the final image on the DSLR sensor. Focus is tuned by translating the objective along the optical axis. The target is mounted with a slant angle of 7.6° to add field depth and further extend the focal dynamic range (FDR) of the setup. Ray colors correspond to different field points on the target. (b) The SHWS design, where a point source is collimated to input a plane wavefront to the LUT. The output wavefront is then measured by a Shack-Hartmann wavefront sensor (SHWS). The figure presents the case where the focal length of the LUT is shorter than the distance between the LUT and SHWS plane. When the focal point is close to the sensor, the resolution is insufficient of measurement causing a 'blind' region. For larger focal lengths (converging beam at the sensor plane), the wavefront is again resolvable. (c) The focal length of a bi-convex spherical lens with a diameter of 25 mm and refractive index of 1.403, as a function of its volume. The first derivative of this curve is also shown to emphasize the sensitivity of the focal length to changes in the volume, which decrease substantially as the volume increases. The gray area marks the unified focal dynamic range covered by the setups. The horizontal white strip represents the small range of focal lengths that cannot be resolved by the SHWS.*

In both designs, the optical elements, and the distances between them dictate the focal dynamic range (FDR) - the range of values for the focal length of the LUT that can be measured. In the case of the slanted target, at least one point on the slanted target must be focused on the sensor within the travel range of the objective. In the case of the SHWS setup, the wavefront slope must stay within a certain range to ensure that the spot centroids fall within the bounds of their corresponding pixel bins[29,30], and the entire beam size must be sufficiently large at the sensor plane to capture it with sufficient resolution. Furthermore, due to the limited numerical aperture of the measurement system, the focal length of the LUT dictates its measurable aperture. To accommodate for the uncertainty in focal length, and since changes to the optical system are not feasible during the short microgravity period, the system must be designed to capture a sufficiently large FDR. This poses a tradeoff as the increase in FDR decreases the active aperture.

In both optical trains we optimized the design to extend the FDR as far as possible without excessively reducing the measured aperture of the LUT. Our final design allowed for an FDR of 47-92 mm with about 30% aperture coverage for the DSLR system, and FDRs of 50-90 mm and 105-400 mm with about 50% aperture coverage for the SHWS system. Both the DSLR and SHWS were configured to work continuously, resulting in frame rates in the range of 5-10 fps.

## 3. Results and Discussion

Figure 3a shows the proper acceleration data recorded by the accelerometer during the flight, with a sampling frequency of 2 Hz. The planned g profile is easily observed – six sets of five parabolas each. The first set was comprised of two 'Martian' (~0.38g) and three 'Lunar' (~0.16g) maneuvers, while the rest were microgravity maneuvers. To analyze the stability of the microgravity environment provided on the aircraft, we ad hoc define 0.1 g as a threshold value under which we consider the conditions to be 'microgravity'. Each microgravity parabola produced a single continuous microgravity phase in which this threshold was not exceeded. The inset in Figure 3b shows a magnified period of one of the maneuvers. Figure 3b presents a histogram showing the relative duration of g values for the union of all microgravity phases in the flight. The histogram shows that during more than 90% of the microgravity time, a clean environment of less than 0.04g is obtained.

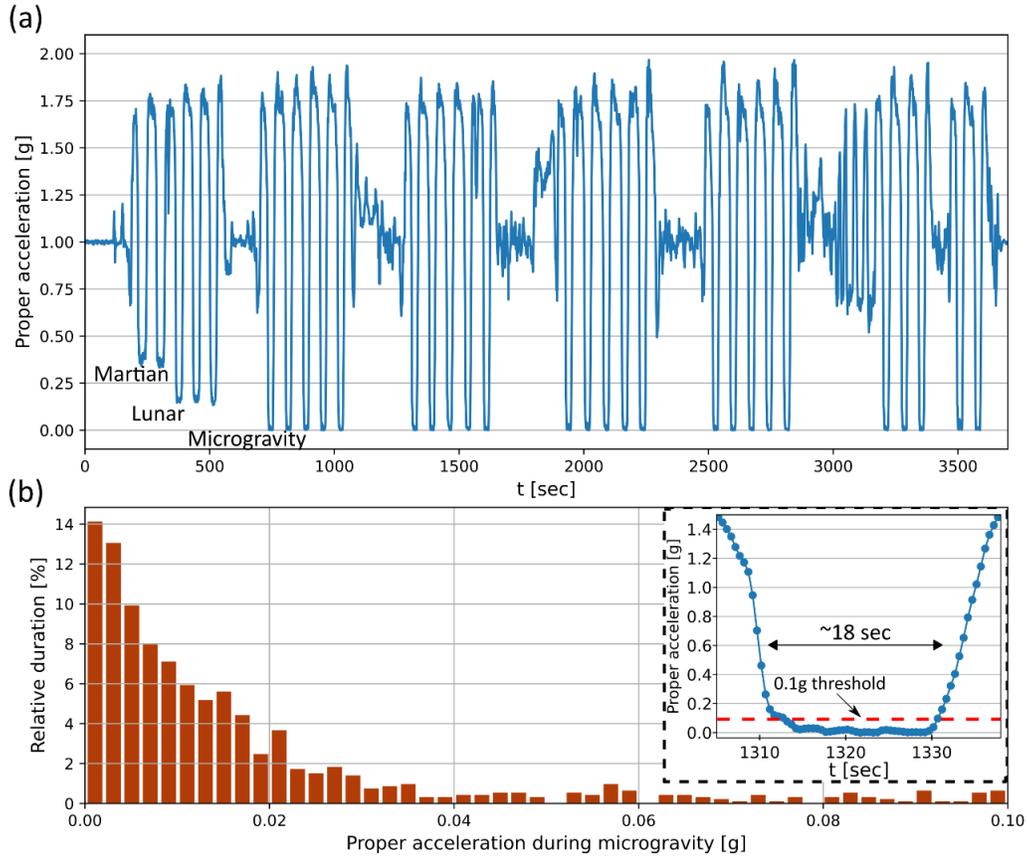

*Figure 3: Acceleration profile during the parabolic flight. (a) The proper acceleration magnitude during the flight as measured by the average of two accelerometers located on the experimental setups. The flight profile is composed of one set of reduced gravity conditions (Martian and Lunar), followed by five sets of microgravity. (b) A histogram showing the relative duration of acceleration magnitude during the microgravity periods (which we ad hoc define as g <=0.1). Over 90% of the microgravity time measures below 0.04g. The inset shows a magnified view of a representative parabolic maneuver in which microgravity conditions were maintained for of 18 seconds.*

Figures 4a-f show several frames from the side camera during the deployment process (see the SI for a video showing such a deployment). Each experiment started with a new lens module containing a pre-filled C-Frame, with black rubber plug sealing its hole (Figure 4a). Upon entering microgravity, we pulled the plug out of the C-Frame hole (Figure 4b) and immediately turned on the pump to initiate the injection of liquid into the frame. In the absence of gravity, the liquid advanced radially from the outer edge of the hole toward its center (Figure 4c-d). When all the liquid fronts met, a continuous liquid lens was formed (Figure 4e). When returning to gravity at the end of each parabola, the liquid lens is drained from the lens hole toward the bottom of the acrylic shield and onto an absorbing pad (Figure 4f).

SI Table 1 provides a table listing all the of the parabolas executed on two flights performed on two consecutive days. On the first flight, out of 14 attempts, 11 resulted in a fully closed lens, as captured by the GoPro camera observing the lens module. Of these, only two were successfully measured (both on the SHWS setup). Between the flight dates, we relaxed the DSLR setup by translating both the LUT and the target 225 mm toward the camera. This results in a wider range of LUT focal lengths that can be captured, at the expense of a smaller measured aperture. The design presented in Figure 2 corresponds to this modified configuration. On the second flight, 12 out of 15 attempts were successful, of which 8 were measured (five by SHWS and three by the DSLR setup). As can be seen in the SI video, even after the lens is fully formed, the shape of the lens fluctuates, either due to injection dynamics that have not fully settled or due to disturbances to the microgravity environment. In some cases, these changes were too rapid or too significant in amplitude to obtain useful measurements. Also, measurements that have been obtained vary significantly due to those changes – typically at a higher rate than the sampling frequency. In the design of the experiment we made a conscious decision to favor yield over measurement statistics, and hence created the lenses using liquids of different viscosities and different injected volumes. There is therefore no clear metric that can be used to compare them, and the sample size is too small to provide valuable statistics. The results we report in the following sections are representative of the better lenses captured in the experiments.

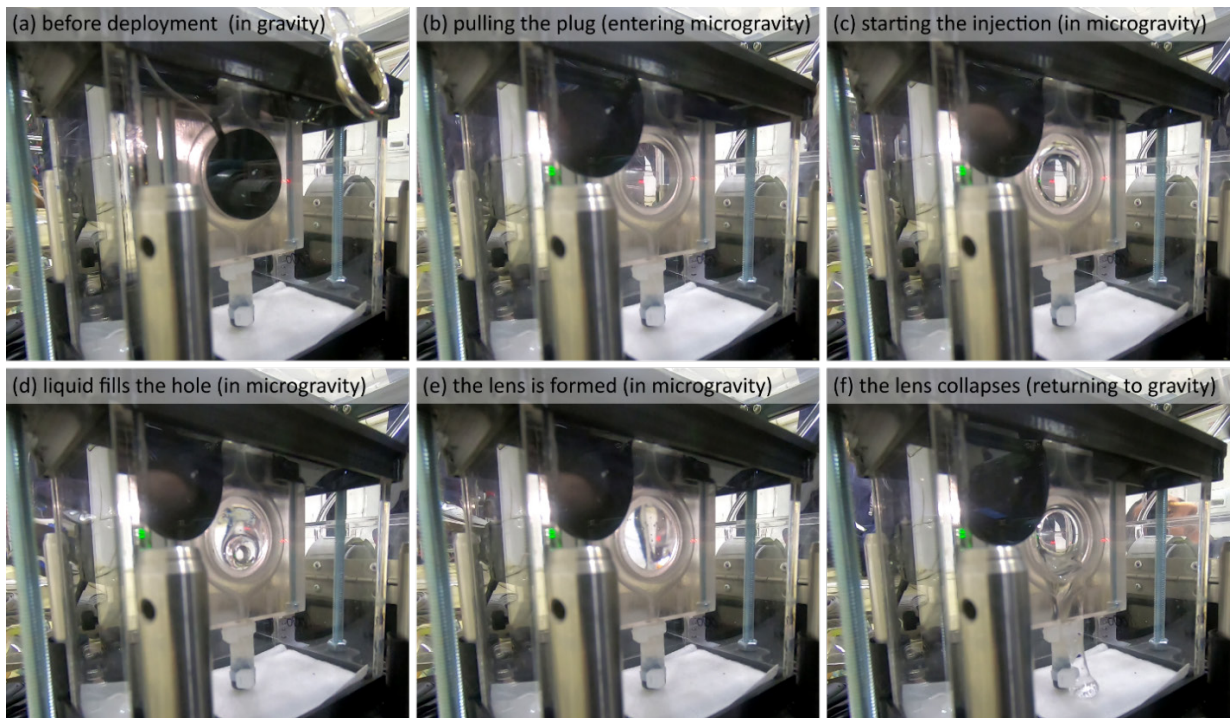

*Figure 4: The deployment of a liquid lens in microgravity, as captured by the GoPro side camera. (a) Before deployment, the rubber plug seals the hole of the C-Frame. (b) Upon entering microgravity, the*

*rubber plug is pulled, and the injection is initiated. (c-d) Liquid flows radially out of the channel and fills the hole until the liquid fronts meet and (e) a liquid lens is formed and is maintained throughout the microgravity duration. (f) Upon returning to gravity, the lens collapses into a puddle at the bottom of the module.*

### 3.1. Direct imaging of a resolution test chart

Figure 5a shows the target that consists of pairs of bright and dark strips (line-space pattern). Each line and its adjacent space are of equal width and constitute a line-pair (lp). Three consecutive pairs with the same width form a group. From left the right, the mask is designed to have 10 groups (with three pairs in each) of decreasing line width from 1 mm to 2 μm. The entire target is slanted at an angle around the horizonal axis, allowing a focused image to be obtained for a wide range of focal lengths of the LUT. Figure 5b-d show the target image as acquired by the DSLR setup, and two consecutive zoomed-in sections of the image presented in grayscale. Figure 5e presents the intensity profile corresponding to Figure 5c, and the resolvable pair groups used for the analysis. For each line-pair group of line width $\delta$, we denote the spatial frequency as $\nu = \frac{1}{2\delta}$, and define the directly measured contrast, $C$, as[31]:

$$C(\nu) = \frac{I_{max}(\nu) - I_{min}(\nu)}{I_{max}(\nu) + I_{min}(\nu)}, \qquad (1)$$

where $I_{max}, I_{min}$ are the max and min grayscale values at spatial frequency $\nu$. In our case, the line pattern is vertical, meaning that we only measure spatial frequencies in the perpendicular direction, i.e., on the sagittal plane of the LUT. Since in our case the target image is binary (sharp steps between 0% and 100% transparency), the directly measured contrast is the square wave modulation, also called the contrast transfer function (CTF). The intrinsic modulation transfer function (MTF) is defined only for spatially monochromatic sine wave targets, but can be obtained directly from the CTF by correcting for the Gibbs phenomena at the edges[32],

$$M(\nu) = \frac{\pi}{4}\left[C(\nu) + \frac{C(3\nu)}{3} - \frac{C(5\nu)}{5} + \frac{C(7\nu)}{7} - \cdots\right] \qquad (2)$$

Since each group consists of three line-pairs, we calculate the CTF and MTF three times for each spatial frequency. Since the Gibbs correction in Equation ( 2 ) requires interpolating the CTF to estimate it in all spatial frequencies, the calculation can be repeated with many permutations of the measured data points. We therefore calculate the MTF separately for the maximum, minimum and intermediate CTF values of each group, leaving a range of values for the MTF at each spatial frequency. Figure 5f presents the resulting MTF range, indicated by the red area. Overall, the MTF shows a gradual decrease in modulation with increasing frequencies, reaching 10% between 40 and 50 lp/mm (2.3-2.8 cycles/mrad based on the estimated focal length of the LUT as described below). The sensor cutoff is at the 60 lp/mm, corresponding to half the sensor's Nyquist frequency, with a pixel size of 4.35 μm. However, we formally indicate the zero modulation at a slightly higher spatial frequency of 64 lp/mm according to our measured pattern. For reference, the green line in Figure 5f presents the simulated MTF of the same optical train, where the LUT is assumed to be a perfectly spherical bi-convex lens with the same focal length and aperture as the measured LUT. This serves as a reference since a spherical lens is the theoretical shape that can be obtained by fluidic shaping in microgravity[25]. Since the focal length and aperture are not directly known from the experiment, we estimate them with a ray tracing simulation where those parameters are adjusted until the measured magnification is obtained at the correct field position (identified by the image position along the target plane). This yields a focal length of $f_{LUT} = 57.1\ mm$ and an aperture of $d_{LUT} = 10.5\ mm$, which fall into their acceptable ranges according to the design. We can see that the measured curve agrees with the ray tracing solution, indicating that within the measured aperture and field position on the optical axis, the LUT in our test setup resolves information similarly to a perfectly spherical lens. We do not imply that the surfaces themselves are perfectly spherical, as multiple aberrations can either balance one another, be less significant in the measured configuration (i.e., on or close to the optical axis), or have no effects on the resolution (e.g., distortion). However, this clearly shows that the LUT is indeed an optical lens that can function within an optical system. For completeness, we provide the MTF of a diffraction limited lens with the same focal length and aperture. Clearly, the MTF of even the ideal spherical lens is far from the diffraction limited one, due the inherent spherical aberration.

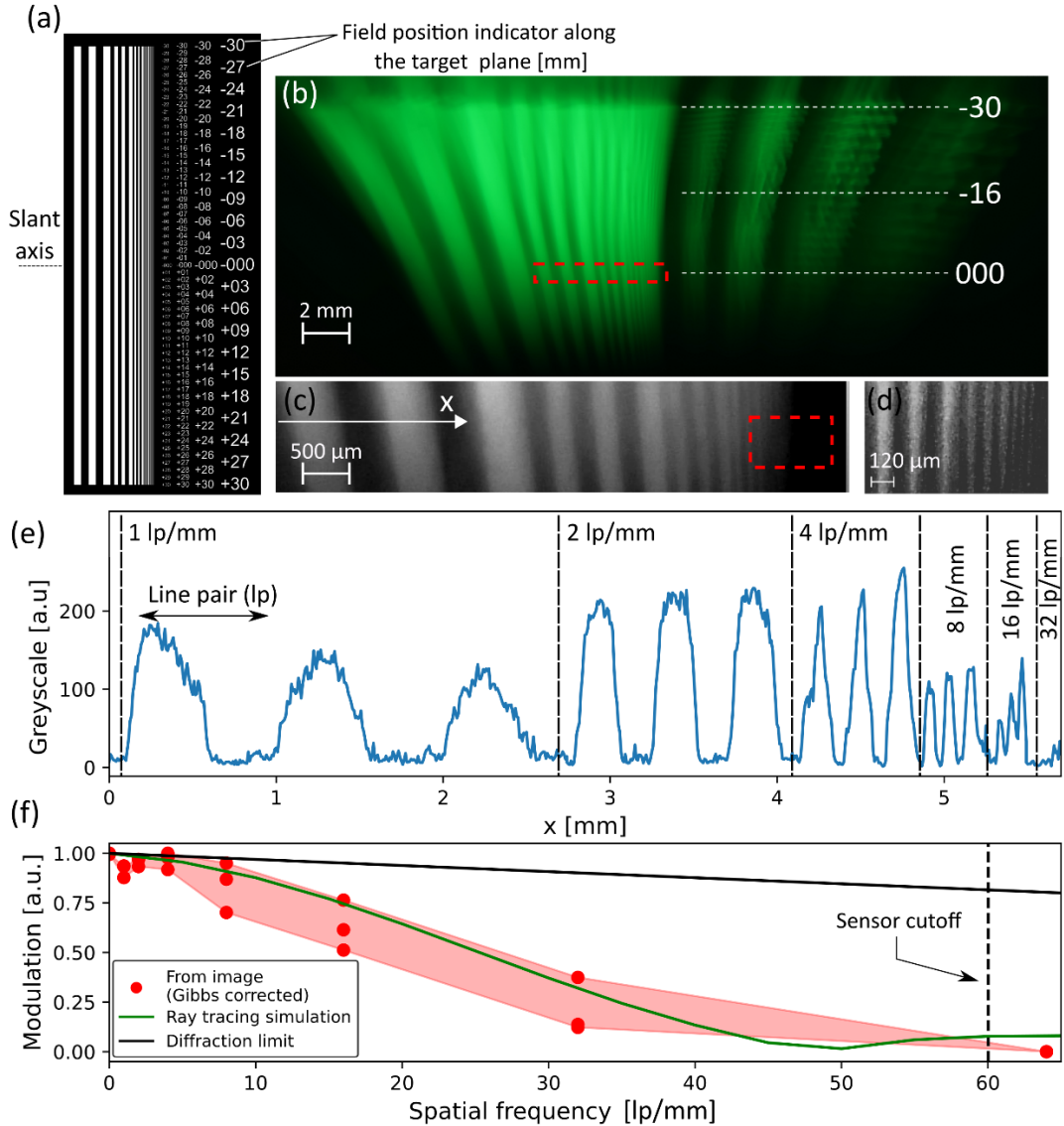

*Figure 5: Evaluation of lens performance through imaging of a slanted resolution-target. (a) The slanted target design containing a vertical line pattern with a variable pitch. The target is slanted around the horizontal axis as marked by the dashed line. The numbers to the right denote position along the target plane, (in mm) and are used to indicate the field position (in mm) of each point in the acquired images. (b-d) An image and two sequential zoomed-in views (marked by red dashed lines) of the slanted target, taken through the LUT in microgravity. (e) The intensity profile taken along the indicated x-axis, and the associated line-pair groups. (f) Comparison of MTF results obtained from analysis of the target image, from ray-tracing simulation of a lens with the same nominal geometry, and from a diffraction limit calculation. The red region corresponds to the range of values that can be extracted from the target image.*

## 3.2. Shack-Hartmann wavefront sensing

Figure 6 presents the wavefront sensing analysis for one of the liquid lenses successfully captured in microgravity. Figure 6a shows the raw measured phase map as it was recorded on the sensor. Figure 6b shows the Zernike decomposition of this phase map, where each vertical bar represents the amplitude of the corresponding Zernike mode coefficient. To estimate the aperture and focal length of the LUT we first calculate the radius of curvature of the wavefront using the Defocus term (with a Zernike coefficient of $Z_5 = -131.1 \ \mu m$)[33,34]. We then find the paraxial focal length and the aperture of the LUT by adjusting these parameters in a ray tracing simulation until the measured wavefront radius of curvature is obtained. After acquiring these values, we discard Piston, Tip, Tilt and Defocus (marked in red in figure 6b) for the rest of the analysis. The main aberration modes observed in Figure 6b are coma and astigmatism. Since the incoming beam is collimated to within a wavefront error of $\lambda/10$ , such asymmetric aberrations suggest corresponding asymmetric deformations on the liquid surfaces, which we suspect to be related to either liquid dynamics from the injection process or non-uniform wetting of the C-Frame surfaces during deployment. Similar deformations can be seen in the SI video, and are common in other deployed lenses as well.   Figure 6c shows the reconstructed phase map by summing all the Zernike terms, excluding Piston, Tip, Tilt and Defocus. The resulting root-mean-squared (RMS) wavefront error is $6.4 \ \mu m$ and represents actual aberrations caused by the LUT. Ideally, the expected geometry of the LUT is a bi-convex spherical lens, which would result in spherical aberration in the form of a quartic radial function[35]. The expected wavefront error RMS of such a lens can be computed by ray tracing to be $3.1 \ \mu m$ (see fig S2 in the SI). The wavefront error phase observed in Figure 6c is clearly not a quartic function and has a much larger RMS error. We therefore conclude that the reconstructed phase in Figure 6(c) is mainly associated with surface irregularities (i.e., 'figure errors') of the LUT, which we thus estimate to be on the order of several microns.

Figures 6d-e present two figures-of-merit for assessing the performance of a lens.  Figure 6d shows the simulated point-spread-function (PSF) representing the distribution of intensity that would be obtained at the focal plane if the LUT is used to focus a collimated beam. A convenient way to compare point spread functions, especially for under-corrected optical systems that are far from the diffraction limit, is by drawing the encircled energy around the centroid of the PSF. In a diffraction limited system, the Airy disk contains 83.8% of the total power[27]. We therefore define the size of a PSF by encircling 83.8% of the power, and present this for both the LUT (1.01 mrad in diameter) and an equivalent spherical lens simulated via ray tracing (0.80 mrad in diameter). Figure 6e presents the two-dimensional MTF obtained from the reconstructed phase, showing a reduction to 10% around 2 cycles/mrad – consistent with the result obtained using target imaging.

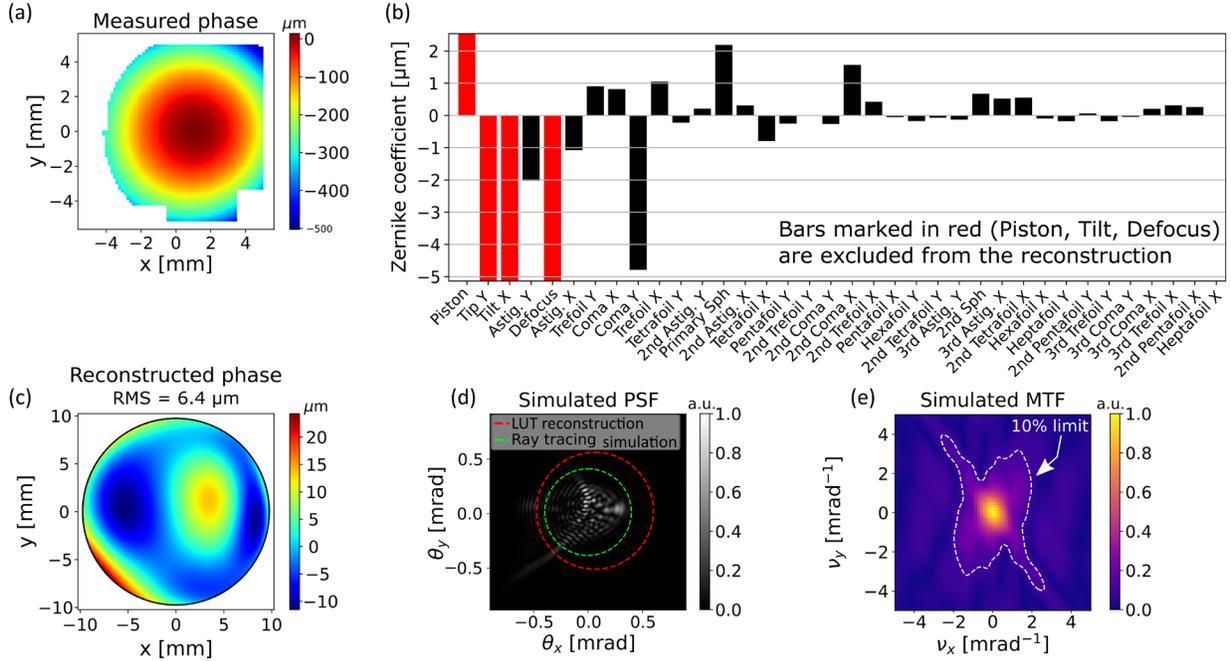

*Figure 6: Evaluation of lens performance through wavefront sensing. (a) The raw data, as obtained by the SHWS, showing the measured phase map at the sensor plane. White pixels correspond to points at which the SHWS could not determine the wavefront slope. (b) The Zernike coefficients of the phase map. The scale is adjusted according to the magnitude of all coefficients except for Piston, Tilt and Defocus (marked in red), as they are excluded from the phase reconstruction. (c) The phase at the lens pupil plane, as reconstructed from the Zernike coefficients in after removing Piston, Tip, Tilt and Defocus (marked in red in 'b'), showing the resulting RMS and P-V of the bare aberrated wavefront. (d) Simulated PSF of the reconstructed phase, after propagating it to the focal plane of the LUT. The red and green circles show the 83.8% encircled energy for the LUT and for an equivalent spherical lens (via ray tracing). (e) Simulated MTF of the reconstructed phase.*

## 4. Conclusions and outlook

In this work, we demonstrated for the first time the feasibility of shaping liquids into lenses under microgravity conditions. Our theory predicts under these conditions the formation of a bi-convex spherical lens, and our target imaging and wavefront sensor measurements indicate optical performance comparable with a spherical lens (as measured by the MTF).

One of the main difficulties in parabolic flight experiments is the time 'budget', i.e., achieving the desired functionality within each parabolic maneuver. Specifically, in our experiments, the challenge was in injecting a very precise volume of liquid in this short period of time. This precision is needed to produce a lens with a focal length that is within the tolerance of the optical system. Our working point was based on open-loop injection which could not account for any unexpected variations in the system, such as due to temperature and pressure changes in the cabin causing expansion or contraction of the liquid. We therefore designed the optical system to accommodate for such variations – this was achieved by the combination of a reduced aperture and the slanted target, both of which increased the focal dynamic range. This approach proved successful, as we were able to measure 10 out of the 23 deployed lenses, across different liquids with a wide range of viscosities and injected volumes. A future improved experimental system could include a feedback loop based on linear encoders of the injection system, on an online flow rate measurement, on the optical measurement, or on any combination of these. This would allow to increase the measured aperture and to better tailor the measurement for the specific LUT.

Our dynamic measurements clearly showed variations in the shape of the lens surfaces during the microgravity periods. We cannot rule out that the reason for this is deviations from 0 g (either static or dynamic), but since our accelerometer measurements indicate a fairly stable microgravity environment, we believe that the main source of variations is the dynamic response of the liquid to the rapid injection. The time 'budget' is an inherent limitation of parabolic flights and introduces an optimization problem that need to be solved – injecting the liquid sufficiently fast (particularly if seeking to create larger apertures) to leave sufficient time for measurement, while minimizing fluctuations after the injection is completed. A closed-loop system would be necessary for this purpose as well. An important part of our experiments was demonstrating the deployment of a liquid lens. However, if one wishes to focus solely on characterizing a liquid lens under microgravity, one possible approach to minimize liquid oscillations would be to skip the deployment process and observe a volume of liquid that nearly fills a horizontal dish and wets its edges, forming a concave lens under microgravity.

Our focus in this work was on lenses, i.e., measuring transparent liquids as refractive elements. Another important category of optics is mirrors, which are useful in space applications due to their spectral range and achromaticity, and lower mass. Thus, in future experiments it would be of interest to consider reflective liquids. An experimental system that is aimed at measuring reflection from a liquid surface would also allow for an easier reconstruction of the surface's shape. This is because each surface would be measured independently, and the measurement would not be affected by volumetric optical properties.

We envision a set of experiments for maturation of the technology toward applications in space manufacturing and telescopes. The limited duration of microgravity conditions during the parabolic flights was sufficient to demonstrate the concept of fluidic shaping under microgravity, but insufficient to allow curing of liquid polymer lenses, even for the fastest reacting polymers that we are aware of. It is also insufficient for complete relaxation of transient effects. Furthermore, deployment of larger aperture optics is not possible within the space and time constraints of such flights. The immediate next step would be to perform experiments onboard the International Space Station (or other orbital flight), providing a longer period of high-quality microgravity. This would enable curing and solidifying high quality lenses over a larger range of apertures and would effectively extend current in-space additive manufacturing capabilities to optics. Another step would be experiments with reflective liquids, demonstrating that the method could be extended to the creation of mirrors. As seen in our experiment, liquid dynamics play an important role in the resulting optical quality. Thus, special attention should be given to the investigation of liquid-structure dynamics, in the context of their effect on the liquid surface shape and optical performance under in-space conditions.

## 5. Acknowledgments

This project has received funding from the European Research Council under the European Union's Horizon 2020 Research and Innovation Programme, grant agreement 10104451 (Fluidic Shaping) and NASA Ames Center Innovation Fund (2020 and 2021). We also greatly acknowledge financial support from Technion's Center for Security Science and Technology (CSST) and from the Norman and Helen Asher Space Research Institute (ASRI) fund. M.E. was supported by the Ramon Graduate Fellowship of the Israel Ministry of Innovation Science and Technology. We thank Aliza Shultzer of Technion and Howard Cannon of NASA Ames for managing the logistics of the parabolic flight experiments. Finally, we would like to express our gratitude to Jill Bauman, Thomas Berndt, Meredith Blasingame, Karen Bradford, Rhys Cheung, Jacob Cohen, Matthew Holtrust, Robert Padilla, and Harry Partridge of NASA Ames, as well as Kent Bress, Judith Carrodeguas, Trenton Roche, Brian Stanford, and Brian Wessel of NASA HQ for enabling and supporting this collaborative effort.

# Supplementary Information

# Fluidic Shaping and *in-situ* Measurement of Liquid Lenses in Microgravity

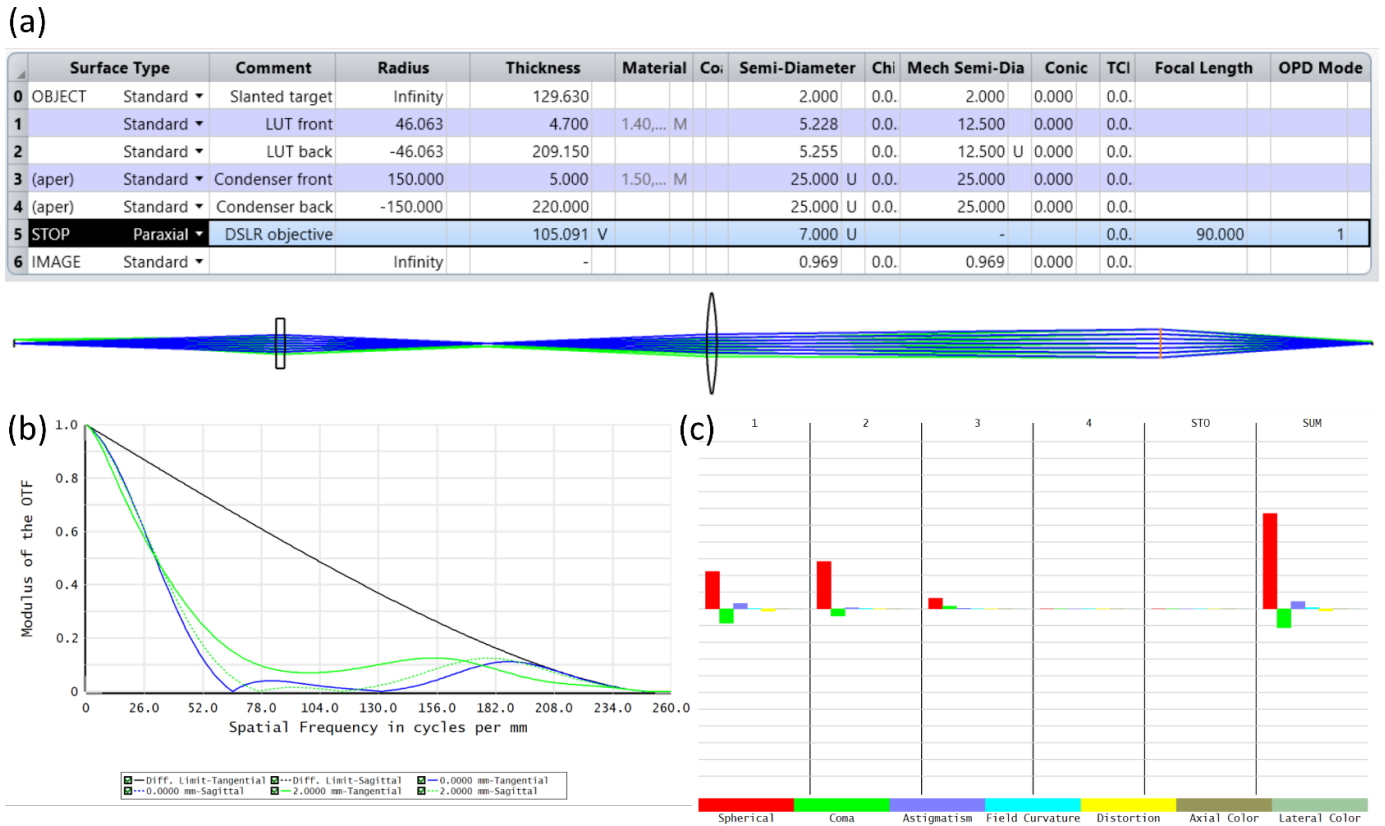

***Figure S1****: Ray tracing simulation of the DSLR setup in Zemax. (a) The lens data table. The design of the Tamron objective is proprietary and therefore, assuming its resolving ability is far better than that of the LUT, we modeled it as a paraxial surface with a 90 mm focal length. (b) The MTF plot of the optical system, showing the sagittal and tangential configurations, compared to that of a diffraction limited system. (c) The Seidel diagram for the optical setup, showing that the majority of aberrations are associated with the surfaces of the LUT.*

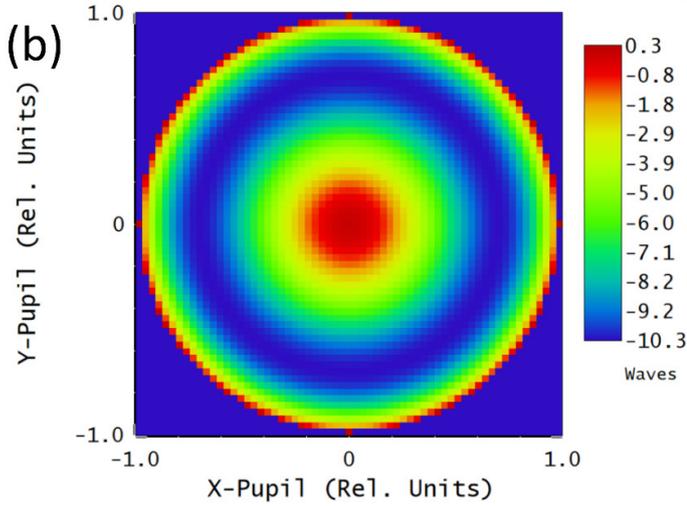
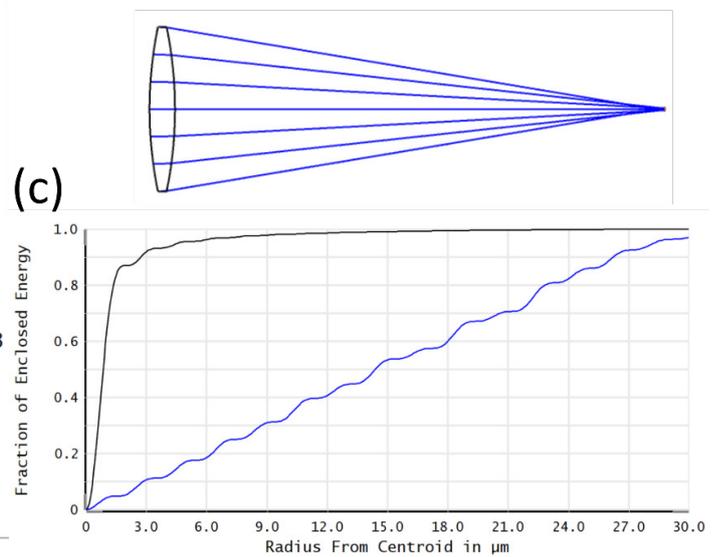

***Figure S2:*** *Ray tracing simulation of the SHWS setup. (a) The lens data table. (b) The associated wavefront, after removing pure defocus. (c) The encircled energy for the case of a spherical LUT versus that of a diffraction limited LUT.*

| Flight date | Parabola set # | Parabola # | g profile | Estimated duration [sec] | Active setup | Viscosity [cSt] | Plug success | Injection success | Deployment success | Data success |
|---|---|---|---|---|---|---|---|---|---|---|
| 09-Dec-21 | 1 | 1 | Martian | 17 | - | - | - | - | - | - |
| 09-Dec-21 | 1 | 2 | Martian | 17 | - | - | - | - | - | - |
| 09-Dec-21 | 1 | 3 | Lunar | 17 | Vertical | 1000 | V | V | V | X |
| 09-Dec-21 | 1 | 4 | Lunar | 17 | SHWS | 1000 | V | V | V | X |
| 09-Dec-21 | 1 | 5 | Lunar | 17 | Slanted | 1000 | V | V | V | X |
| 09-Dec-21 | 2 | 6 | 0g | 14 | Slanted | 1000 | V | V | X | X |
| 09-Dec-21 | 2 | 7 | 0g | 18 | Vertical | 1000 | V | V | V | X |
| 09-Dec-21 | 2 | 8 | 0g | 14 | SHWS | 1000 | V | V | X | X |
| 09-Dec-21 | 2 | 9 | 0g | 18 | - | - | - | - | - | - |
| 09-Dec-21 | 2 | 10 | 0g | 16 | - | - | - | - | - | - |
| 09-Dec-21 | 3 | 11 | 0g | 16 | Vertical | 5000 | V | V | V | X |
| 09-Dec-21 | 3 | 12 | 0g | 14 | SHWS | 5000 | V | V | X | X |
| 09-Dec-21 | 3 | 13 | 0g | 14 | Slanted | 5000 | X | X | X | X |
| 09-Dec-21 | 3 | 14 | 0g | 12 | Slanted | 5000 | V | V | V | X |
| 09-Dec-21 | 3 | 15 | 0g | 14 | - | - | - | - | - | - |
| 09-Dec-21 | 4 | 16 | 0g | 14 | Vertical | 5000 | V | V | V | X |
| 09-Dec-21 | 4 | 17 | 0g | 14 | SHWS | 5000 | V | V | V | V |
| 09-Dec-21 | 4 | 18 | 0g | 14 | Slanted | 5000 | V | V | V | X |
| 09-Dec-21 | 4 | 19 | 0g | 14 | - | - | - | - | - | - |
| 09-Dec-21 | 4 | 20 | 0g | 14 | - | - | - | - | - | - |
| 09-Dec-21 | 5 | 21 | 0g | 15 | Slanted | 1000 | V | V | V | X |
| 09-Dec-21 | 5 | 22 | 0g | 16 | SHWS | 1000 | V | V | V | V |
| 09-Dec-21 | 5 | 23 | 0g | 14 | Vertical | 1000 | V | V | V | X |
| 09-Dec-21 | 5 | 24 | 0g | 14 | - | - | - | - | - | - |
| 09-Dec-21 | 5 | 25 | 0g | 15 | - | - | - | - | - | - |
| 09-Dec-21 | 6 | 26 | 0g | 13 | Slanted | 1000 | V | V | V | X |
| 09-Dec-21 | 6 | 27 | 0g | 14 | SHWS | 1000 | X | X | X | X |
| 09-Dec-21 | 6 | 28 | 0g | 13 | Vertical | 1000 | V | V | V | X |
| 09-Dec-21 | 6 | 29 | 0g | 14 | - | - | - | - | - | - |
| 09-Dec-21 | 6 | 30 | 0g | 14 | - | - | - | - | - | - |
| 10-Dec-21 | 1 | 1 | Martian | | - | - | - | - | - | - |
| 10-Dec-21 | 1 | 2 | Martian | | - | - | - | - | - | - |
| 10-Dec-21 | 1 | 3 | Lunar | 18 | Vertical | 1000 | V | V | X | X |
| 10-Dec-21 | 1 | 4 | Lunar | 20 | SHWS | 1000 | V | V | V | V |
| 10-Dec-21 | 1 | 5 | Lunar | ? | Slanted | 1000 | V | V | X | X |
| 10-Dec-21 | 2 | 6 | 0g | 14 | Vertical | 1000 | V | V | V | X |
| 10-Dec-21 | 2 | 7 | 0g | 14 | SHWS | 1000 | V | V | V | V |
| 10-Dec-21 | 2 | 8 | 0g | ? | Slanted | 1000 | V | V | V | V |
| 10-Dec-21 | 2 | 9 | 0g | ? | - | - | - | - | - | - |
| 10-Dec-21 | 2 | 10 | 0g | ? | - | - | - | - | - | - |

| Date | | | | | | | | | | | |
|---|---|---|---|---|---|---|---|---|---|---|---|
| 10-Dec-21 | 3 | 11 | 0g | 13 | SHWS | 1000 | V | V | V | V |
| 10-Dec-21 | 3 | 12 | 0g | 13 | Vertical | 1000 | V | V | V | X |
| 10-Dec-21 | 3 | 13 | 0g | 13 | Slanted | 1000 | V | V | V | V |
| 10-Dec-21 | 3 | 14 | 0g | 12 | - | - | - | - | - | - |
| 10-Dec-21 | 3 | 15 | 0g | 13 | - | - | - | - | - | - |
| 10-Dec-21 | 4 | 16 | 0g | 13 | SHWS | 5000 | V | V | V | V |
| 10-Dec-21 | 4 | 17 | 0g | 14 | Vertical | 5000 | V | V | X | X |
| 10-Dec-21 | 4 | 18 | 0g | 13 | Slanted | 5000 | V | V | V | V |
| 10-Dec-21 | 4 | 19 | 0g | 18 | - | - | - | - | - | - |
| 10-Dec-21 | 4 | 20 | 0g | 17 | - | - | - | - | - | - |
| 10-Dec-21 | 5 | 21 | 0g | 14 | SHWS | 5000 | V | V | V | V |
| 10-Dec-21 | 5 | 22 | 0g | ? | Vertical | 5000 | V | V | V | X |
| 10-Dec-21 | 5 | 23 | 0g | 14 | Slanted | 5000 | V | V | V | |
| 10-Dec-21 | 5 | 24 | 0g | 15 | - | - | - | - | - | - |
| 10-Dec-21 | 5 | 25 | 0g | 15 | - | - | - | - | - | - |
| 10-Dec-21 | 6 | 26 | 0g | 15 | SHWS | 200 | V | V | V | |
| 10-Dec-21 | 6 | 27 | 0g | 17 | Vertical | 200 | X | X | X | X |
| 10-Dec-21 | 6 | 28 | 0g | 13 | Slanted | 200 | | | | |
| 10-Dec-21 | 7 | 29 | 0g | 14 | Vertical | 200 | V | V | X | X |
| 10-Dec-21 | 7 | 30 | 0g | 15 | - | - | - | - | - | - |